\documentclass[preprint,12pt]{elsarticle}

\usepackage{graphicx}
\usepackage[caption=false]{subfig}	%for subfloats, option for clash with captions in documentclass
\usepackage{amssymb}
\usepackage{amsmath}
\usepackage{mathrsfs}
\usepackage{dsfont}
\usepackage{slashed}

\usepackage{ulem} %cross out text

\usepackage{color}

\usepackage{vmargin}
\setmargins  {2.5cm}{1.5cm}% % linker & oberer Rand
             {15cm}{22.5cm}%   % Textbreite und -hoehe
             {28pt}{15pt}%   % Kopfzeilenhoehe und -abstand
             {0pt}{40pt}%    % \footheight (egal) und Fusszeilenabstand

\begin{document}

\begin{frontmatter}



\title{\hfill \normalsize{}\vspace{1cm}\\
\Large A lattice study of a chirally invariant Higgs-Yukawa model including 
a higher dimensional $\Phi^6$-term}


\author{David Y.-J.\ Chu\fnref{a}}
\author{Karl Jansen\fnref{b}}
\author{Bastian Knippschild\fnref{c}} 
\author{C.-J.\ David Lin\fnref{d}}
\author{Attila Nagy \fnref{b,e}}

\address[a]{Department of Electrophysics, National Chiao-Tung University, Hsinchu 30010, Taiwan}
\address[b]{NIC, DESY, Platanenallee 6, D-15738 Zeuthen, Germany }
\address[c]{HISKP, Nussallee 14-16, D-53115 Bonn, Germany }
\address[d]{Institute of Physics, National Chiao-Tung University, Hsinchu 30010, Taiwan}
\address[e]{Humboldt-Universit\"at zu Berlin, Institut f\"ur Physik, Newtonstr. 15, \par D-12489 Berlin, Germany }

\begin{abstract}
We discuss the non-thermal phase structure of a chirally invariant Higgs-Yukawa model 
on the lattice in the presence of a higher dimensional $\Phi^6$-term. 
For the exploration of the phase diagram we 
use analytical, lattice perturbative calculations of the constraint
effectice potential as well as numerical simulations.  
We also present first results of the effects of the $\Phi^6$-term
on the lower Higgs boson mass bounds. 
\end{abstract}


\end{frontmatter}

\newpage

\section{Introduction}
\label{sec:intro}

In this letter we   
investigate the influence of the addition of
a dimension-6 operator to a chirally invariant Higgs-Yukawa model.
This model can be understood as a limit of the standard model (SM) without gauge fields.
In particular, we 
consider a complex scalar doublet and one doublet of mass-degenerate quarks.
Our aim is to explore, whether a dimension-6 operator, 
for which we will employ a $(\varphi^{\dagger} \varphi)^3$-term with a
coupling constant $\lambda_6$, can 
modify the phase structure of the Higgs-Yukawa sector of the 
SM and may alter the lower Higgs boson mass bound as already 
observed in \cite{Gies:2013fua,Gies:2014xha}. 
For a phenomenological analysis of a $(\varphi^{\dagger} \varphi)^3$-term 
see e.g.~\cite{Ellis:2014dva,Biekoetter:2014jwa}.

The motivation for adding a $(\varphi^{\dagger} \varphi)^3$-term is
twofold. First, since the Higgs-Yukawa sector of the SM
is trivial, the cut-off cannot be removed and hence such a
term is in principle allowed. In addition, if small values of the cut-off of
$O(1) -- O(10){\rm TeV}$ are considered as done in this work, such a term
can have a significant effect.
Second, the apperance of a $(\varphi^{\dagger} \varphi)^3$-term can be
understood to arise from an extension of the SM. Studying the
system with such a term could hence provide bounds on the couplings
of such extensions in case the lower Higgs boson mass bound 
is incompatible with the Higgs boson mass
of about 126GeV. The effects of higher dimensional operators on the vacuum stability 
is discussed in~\cite{Branchina:2013jra,Branchina:2014usa,Eichhorn:2015kea} .

We use a lattice regularization of the Higgs-Yukawa model which 
eventually also allows non-perturbative numerical simulations for 
large values of $\lambda_6$.
The notion of an exact lattice chiral symmetry \cite{Luscher:1998pqa} 
which derives from the Ginsparg-Wilson relation \cite{Ginsparg:1981bj}
allows us to emulate the continuum Higgs-Yukawa sector of the standard 
model on a discrete Euclidean space-time lattice. To this end, 
the overlap operator \cite{Neuberger:1997fp,Neuberger:1998wv} as 
a local \cite{Hernandez:1998et} lattice Dirac operator has been employed to study the 
phase structure of the lattice theory \cite{Gerhold:2007yb,Gerhold:2007gx}, 
to derive lower and upper Higgs boson mass 
bounds \cite{Gerhold:2009ub,Gerhold:2010bh,Gerhold:2010wv,Bulava:2013ep}
and to analyze the Higgs boson resonance non-perturbatively \cite{Gerhold:2011mx}. 
For a review, see \cite{Bulava:2012rb}. 

For our investigations 
we perform analytical calculations of the phase structure
of the model by computing the constraint effective 
potential (CEP) \cite{O'Raifeartaigh:1986hi} to the first non-trivial 
order in lattice perturbation theory. 
In this calculation, we employ 
the same chirally invariant lattice formulation of the Higgs-Yukawa model as it is 
used for the numerical computations. 
We compare results for the phase structure obtained from numerical simulations 
to our perturbative predictions. 
In addition, we will provide first results for the lower Higgs boson mass 
bounds in the presence of the dimension-6 operator 
as obtained from the analytical, perturbative calculations 
of the CEP.

\section{Basic definitions}
\label{sec:def}

In this work, we restrict ourselves to the case of one fermion 
doublet $\psi=(t, b)^T$ with mass degenerate quarks. The scalar 
fields are a complex doublet $\varphi$. 
Here, we will only provide the basic definitions of the model and refer
to ref.~\cite{Bulava:2012rb} for a more detailed explanantion. 
In Euclidean space time the continuum action is given by:
\begin{multline}\label{eq:action_continuum}
 S^{\text{cont}}[\bar{\psi}, \psi, \varphi] = \int d^4 x \left\{\frac{1}{2}\left(\partial_{\mu} \varphi \right)^{\dagger} \left(\partial^{\mu} \varphi \right)
			+  \frac{1}{2} m_0^2 \varphi^{\dagger} \varphi 
			+ \lambda \left(\varphi^{\dagger} \varphi \right)^2 
			+ \lambda_6 \left(\varphi^{\dagger} \varphi \right)^3 \right\} \\
			+\int d^4 x  \left\{\bar{t} \slashed \partial t + \bar{b} \slashed \partial b +
			y \left( \bar{\psi}_{_L} \varphi\, {b}_{_R} + \bar{\psi}_{_L} \tilde \varphi\, {t}_{_R} \right)
			+ h.c. \right\},
\end{multline}
with $\tilde\varphi = i\tau_2\varphi^*$ and $\tau_2$ being the second Pauli matrix. Besides the standard bare parameters 
$m_0^2$ and $\lambda$ for the Higgs potential and $y$ for the Yukawa coupling, 
we add the dimension-6 operator $\lambda_6 \left(\varphi^{\dagger} \varphi \right)^3$ 
to the action. 

For the numerical implementation of this model we use a polynomial 
hybrid Monte Carlo algorithm\cite{Frezzotti:1998eu} with dynamical overlap 
fermions, see ref.~\cite{Gerhold:2010wy} for details.
On the lattice, it is convenient to rewrite the bosonic part of the action 
in the following way\footnote{The lattice spacing is set to one throughout this paper.}:
\begin{equation}\label{eq:bosonic_action_lattice}
S_B[\Phi] = -\kappa \sum\limits_{x,\mu} \Phi_x^{\dagger} \left[\Phi_{x+\mu} + \Phi_{x-\mu}\right] + 
					\sum\limits_{x} \left( 
					\Phi_x^{\dagger} \Phi_x + 
					\hat{\lambda} \left[ \Phi_x^{\dagger} \Phi_x - 1 \right]^2 + 
					\hat{\lambda}_6 \left[ \Phi_x^{\dagger} \Phi_x \right]^3 \right).
\end{equation}
Here the scalar field, $\Phi$, is represented as a real four-vector and the 
relation to the continuum notation is given by:
\begin{equation}\label{eq:rescaling_cont_lattice}
 \varphi = \sqrt{2 \kappa} \left( \begin{array}{c} \Phi^2 + i\Phi^1 \\ \Phi^0 - i \Phi^3 \end{array} \right) ,\quad
	 m_0^2 = \frac{1 - 2 \hat{\lambda} -8 \kappa}{\kappa} ,\quad
	 \lambda = \frac{\hat{\lambda}}{{4 \kappa^2}},\quad
	 \lambda_6 = \frac{\hat{\lambda}_6}{{8 \kappa^3}}.
\end{equation}
As said above, our main goal is the exploration of the 
phase structure of the model in the presence of the $ \left[ \Phi_x^{\dagger} \Phi_x \right]^3$ 
term with coupling strength $\lambda_6$. We will use 
the magnetization $m$ as the order parameter\footnote{Here we are only interested
in transitions between the symmetric and the spontaneously broken 
phases and thus will not consider the staggered magnetization \cite{Gerhold:2007yb,Gerhold:2007gx}.}. 
The magnetization is given by the
modulus of the average scalar field and is related to the 
vacuum expectation value ($vev$) via:
\begin{equation}\label{eq:mag_and_vev}
 m = \left< \left | \frac{1}{V}\sum\limits_x \Phi_x \right| \right>,\qquad
 vev = \sqrt{2 \kappa} \cdot m.
\end{equation}

For a determination and detailed discussion of the phase structure 
of the model for $\lambda_6=0$, we refer to 
refs.~\cite{Gerhold:2007yb,Gerhold:2007gx}. 

\section{The constraint effective potential}
\label{sec:CEP}
Before resorting to numerical simulations, we study 
the phase structure analytically in lattice 
perturbation theory for which we 
employ the CEP \cite{Fukuda:1974ey, O'Raifeartaigh:1986hi}. 
We assume the scalar field to be in the broken phase, so the scalar field decomposes into 
the Higgs mode, $h$, and the three Goldstone modes, $g^{\alpha}$, with $\alpha=1,2,3$.
The CEP $U(\hat v)$ is described by the  
zero mode of the Higgs field, $\tilde h_0 = V^{-1/2} \hat v$.
The perturbative calculations are done by keeping the lattice 
regularization explicitly, i.e.\ the overlap operator 
is used for the fermionic contribution and all sums over lattice momenta are performed numerically.

To obtain the potential the bosonic non-zero modes are integrated out. To do so, the bosonic 
action is separated into a Gaussian contribution which can be integrated out leading to the bosonic propagators. The remaining terms are treated as an 
interaction part and can be expanded in powers of the couplings. This separation into a Gaussian and an interaction part however is not unique and we 
employ two versions of the CEP.

A derivation of such a lattice constrained effective potential  
can be found in \cite{Gerhold:2007gx,Gerhold:2010wy}. Following the procedure in these references, the Gaussian contribution to the action reads:
\begin{equation}\label{eq:bosonic_gauss_const}
 S^{\text{gauss}}_1[h,g^{\alpha}] = \frac{1}{2}\sum\limits_{p \neq 0} \left( 
					\tilde h_{-p} \left( \hat p^2 + m_0^2 \right) h_p + 
					\sum\limits_{\alpha} \tilde g^{\alpha}_{-p} \left( \hat p^2 + m_0^2 \right) \tilde g^{\alpha}_p \right),
\end{equation}
which leads to the propagator sums:
\begin{equation} \label{eq:def_propagator_sums_bare}
 P_{H} = P_{G} = \frac{1}{V} \sum\limits_{p \neq 0} \frac{1}{{\hat p}^2 + m_{0}^2}.
\end{equation}
As in \cite{Gerhold:2009ub}, $m_0^2$  is replaced by the renormalized masses in the propagator sums. 
The the mass of the Goldstone boson is set explicitly to zero. This leads to:
% The propagator sums for the Higgs and the Goldstone bosons are given by:
\begin{equation} \label{eq:def_propagator_sums}
 P_{H} = \frac{1}{V} \sum\limits_{p \neq 0} \frac{1}{{\hat p}^2 + m_{H}^2} ,\qquad
 P_{G} = \frac{1}{V} \sum\limits_{p \neq 0} \frac{1}{{\hat p}^2}.
\end{equation}
The determinant from integrating out eq.~\eqref{eq:bosonic_gauss_const} is independent of $\hat v$ and can therefore be neglected for the 
CEP.

The CEP up to the first order in $\lambda$ and $\lambda_6$ is then given by:
\begin{multline}\label{eq:CEP_with_phi_6}
 U_1(\hat v) = U_f(\hat v) + \frac{m_0^2}{2} {\hat v}^2 +\lambda {\hat v}^4 + \lambda_6 {\hat v}^6 \\
                         + \lambda \cdot {\hat v}^2 \cdot 6(P_H+P_G)
                         + \lambda_6 \cdot \left( {\hat v}^2 \cdot ( 45 P_H^2 + 54 P_G P_H + 45 P_G^2)
                         + {\hat v}^4 \cdot ( 15 P_H + 9 P_G ) \right ).
\end{multline}
The fermionic contribution, $U_f$, originates from integrating out the 
fermions in the background of a constant field. It takes the form,
\begin{equation} \label{eq:fermionic_contribution_CEP_massbound}
 U_f(\hat v) = -\frac{4}{V} \sum\limits_p \log\left| \nu^+(p) + y \cdot \hat v \cdot \left( 1-\frac{\nu^+(p)}{2 \rho} \right)  \right|^2,
\end{equation}
where $\nu^{\pm}(p)$ denotes the eigenvalues of the overlap operator,
\begin{equation} \label{eq:eigenvalues_of_overlap}
 \nu^{\pm}(p) = \rho \left( 1 + \frac{ \pm i \sqrt{{\tilde p} ^2} + r  {\hat p}^2 - \rho}
                           {\sqrt{ {\tilde p} ^2 + \left( r  {\hat p}^2 - \rho\right)^2}}\right),\quad
 {\hat p}^2 = 4 \sum\limits_{\mu} \sin^2\left(\frac{p_{\mu}}{2}\right),\quad
 {\tilde p}^2 = \sum\limits_{\mu} \sin^2\left(p_{\mu}\right).
\end{equation}
In this equation $r$ denotes the Wilson parameter and $\rho$ ($0 \leq \rho \leq 2r$) is a free 
parameter of the overlap operator which can be tuned to optimize its 
locality properties \cite{Hernandez:1998et}.
Throughout this work, we set $r=1$ and $\rho=1$. 

In addition to the procedure leading to $U_1(\hat v)$, eq.~\eqref{eq:CEP_with_phi_6}, 
another ansatz in performing the Gaussian integral is to collect all the terms that are quadratic in the bosonic non-zero modes from the 
self interaction:
\begin{multline}\label{eq:bosonic_gauss_with_zero_mode}
 S^{\text{gauss}}_2[h,g^{\alpha}] = \frac{1}{2}\sum\limits_{p \neq 0} \left( 
			\tilde h_{-p} \left( \hat p^2 + m_0^2 + 12 \lambda \hat v ^2 + 30 \lambda_6 \hat v^4 \right) \tilde h_{p} \right. \\
			\left. + \sum\limits_{\alpha} \tilde g^{\alpha}_{-p} \left( \hat p^2 + m_0^2 + 4 \lambda \hat v^2 + 6 \lambda_6 \hat v ^4 \right) 
			\tilde g^{\alpha}_p \right).
\end{multline}
In this approach the bosonic determinant can no longer be neglected for in potential calculation, since it
depends explicitly on the zero mode.  
Further, at first order in $\lambda$ and $\lambda_6$ of perturbation theory, the propagator sums and combinatorial factors 
change,
\begin{align}\label{eq:CEP_with_phi_6_withFullBosDet}
 U_2(\hat v) & =  U_f(\hat v) + \frac{m_0^2}{2} {\hat v}^2 +\lambda {\hat v}^4 + \lambda_6 {\hat v}^6 
        \nonumber \\
             & + \frac{1}{2 V} \sum\limits_{p \neq 0} 
                 \log \left[  \left( \hat p^2 + m_0^2 + 12 \lambda \hat v^2 + 30 \lambda_6 \hat v^4 \right) \cdot  
                 \left( \hat p^2 + m_0^2 + 4 \lambda \hat v^2 + 6 \lambda_6 \hat v^4 \right)^3  \right]
         \nonumber \\
             & + \lambda \left ( 3 \, \tilde P_H^2 + 6 \, \tilde P_H \tilde P_G + 15 \, \tilde P_G^2 \right)   
                 + \lambda_6 \hat v^2 \left ( 45 \, \tilde P_H^2 + 54 \, \tilde P_H \tilde P_G + 45 \, \tilde P_G^2 \right) 
         \nonumber \\
             &  + \lambda_6 \left( 15 \, \tilde P_H^3 + 27 \, \tilde P_H^2 \tilde P_G + 45 \, \tilde P_H \tilde P_G^2 + 105 \, \tilde P_G^3 \right),
\end{align}
with the propagator sums given by:
\begin{equation}\label{eq:propagatorSum_withFullBosDet}
\tilde P_H = \frac{1}{V} \sum\limits_{p\neq 0} \frac{1}{ \hat p^2 + m_0^2 + 12 \hat v^2 \lambda + 30 \hat v^4 \lambda_6 }, \quad
\tilde P_G = \frac{1}{V} \sum\limits_{p\neq 0} \frac{1}{ \hat p^2 + m_0^2 + 4 \hat v^2 \lambda + 6 \hat v^4 \lambda_6 }.
\end{equation}

In this approach logarithmic terms appear. 
Depending on the choice of the bare parameters $(m_0^2, \lambda, \lambda_6)$, 
the arguments of the logarithms may become negative, 
leading to the well known problem that the effective potential becomes
complex \cite{Weinberg:1987vp}. 
We remind, that the lattice spacing is set to one implicitly such 
that, even though we use the continuum notation, 
all quantities are dimensionless.

Using the analytical form of the CEP, 
the $vev$ can be obtained by the (absolute) minimum of the potential. 
In order to introduce a physical scale, we set the lattice $vev$
to the phenomenologically known value of 246~GeV and define the cutoff, $\Lambda$ as the inverse lattice spacing:
\begin{equation}\label{eq:vev_and_cutoff_CEP}
 \left. \frac{\text{d}U(\hat v)}{\text{d}\hat v} \right|_{\hat v = vev} \stackrel{!}{=} 0 ,\qquad
 \Lambda = \frac{246 \text{ GeV}}{vev}.
\end{equation}
Further, the squared Higgs boson mass $m_H^2$ is determined by the 
second derivative of the potential at its minimum,
\begin{equation}\label{eq:mhSquared_from_CEP}
 \left. \frac{\text{d}^2 U(\hat v)}{\text{d}{\hat v}^2} \right|_{\hat v = vev} = m_H^2.
\end{equation}
Due to the explicit appearance of the Higgs boson mass in the 
propagator sum eq.~\eqref{eq:def_propagator_sums} for the potential $U_1$, 
eq.~\eqref{eq:CEP_with_phi_6}, we have to use an 
iterative approach in the determination of
a solution for the minimum of the CEP and the Higgs boson mass. To this end, we fix the parameters 
$m_0^2$, $y$, $\lambda$ and $\lambda_6$, guess an initial Higgs boson mass
and simply iterate  eqs.~(\ref{eq:vev_and_cutoff_CEP},\ref{eq:mhSquared_from_CEP}) until
we find convergence.

We will compare results obtained from both forms of the potential to results from our non-perturbative simulations. 
As we will see below, we indeed find parameter sets, where the perturbative CEP describes 
the non-perturbative data well, even on a quantitative level. 
This will allow us 
to obtain results for the phase structure of the Higgs-Yukawa model considered here   
from the analytical perturbative CEP, where a non-perturbative simulation is 
not feasible anymore, i.e. for large lattices or large cut-offs.

\section{Results}
\label{sec:results}

For our study of the phase structure we performed simulations for two
values of $\lambda_6$ (0.001 and 0.1).
Note, that having set the lattice spacing to 
one, $\lambda_6$ is treated as a dimensionless coupling constant.  
For each value of $\lambda_6$ we choose a set of values for 
the quartic coupling, $\lambda$. 
The Yukawa coupling, $y$, is chosen such that the quarks in our model have a mass of that of the physical top quark, 
$m_t = y \cdot vev \cdot \Lambda\approx 175$~GeV. The phase transition
between the symmetric and spontaneously broken phases is probed by scanning
in the hopping parameter, $\kappa$.

%%%%%%%%%%%%%%%%%%%%%%%%%%%%%%%%%%%%%%%%%%%%%%%%%%%%%%%%%%%%%%%%%%%%%%
%%   vev vs kappa l6=0.001 and 0.1, various lambda
%%%%%%%%%%%%%%%%%%%%%%%%%%%%%%%%%%%%%%%%%%%%%%%%%%%%%%%%%%%%%%%%%%%%%%
\begin{figure}[htb]
\centering
\subfloat[$\lambda_6=0.001$]{\includegraphics[width=0.5\linewidth]
{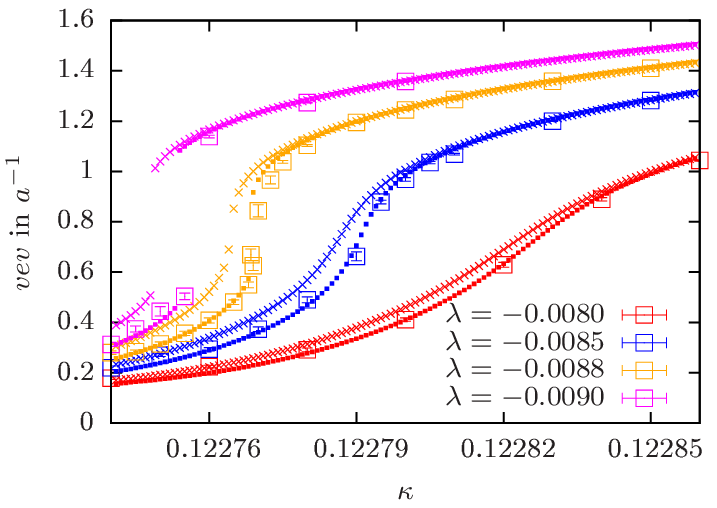}\label{fig:vev_vs_kappa_l6_0.001}}
\subfloat[$\lambda_6=0.1$]{\includegraphics[width=0.5\linewidth]
{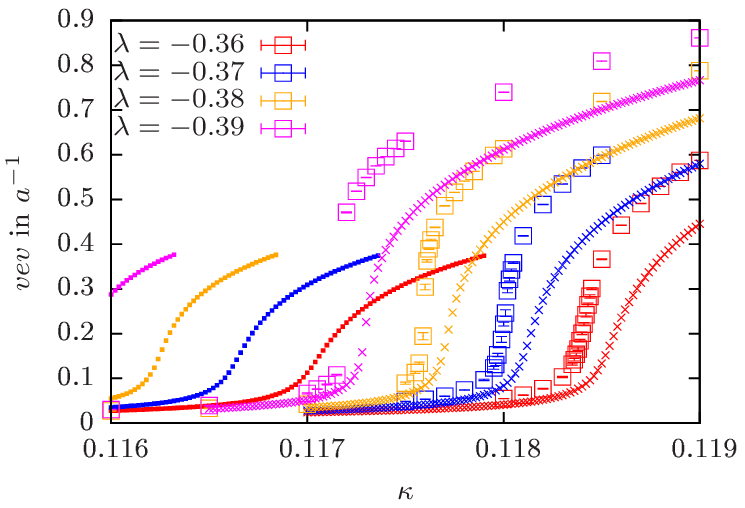}\label{fig:vev_vs_kappa_l6_0.1}}
\caption{
% Here we show a comparison between 
Data which were obtained from numerical simulations and the perturbative 
approaches described in section~\ref{sec:CEP} are compared. 
The plots show the $vev$ as a function of $\kappa$ while $\lambda_6$ is kept fixed to 
$\lambda_6=0.001$ (left) and $\lambda_6=0.1$ (right) for various $\lambda$. 
The simulation data are depicted by the open squares, 
the crosses indicate the $vev$ obtained from
$U_1$, eq.~\eqref{eq:CEP_with_phi_6}, while the dots show the corresponding results from 
$U_2$, eq.~\eqref{eq:CEP_with_phi_6_withFullBosDet}. All data have been obtained
on $16^3 \times 32$ lattices.}
\label{fig:vev_vs_kappa}
\end{figure}
%%%%%%%%%%%%%%%%%%%%%%%%%%%%%%%%%%%
%%%%%%%%%%%%%%%%%%%%%%%%%%%%%%%%%%%

In figure~\ref{fig:vev_vs_kappa} we show results for the bare $vev$ 
computed on lattices with volume $16^3 \times 32$ 
for $\lambda_6=0.001$ (left) and $\lambda_6=0.1$ (right).
Our data show the same qualitative behaviour for both values of $\lambda_6$. 
The phase transition is of second order when $\lambda$ is chosen negative and its absolute value is small. Increasing the absolute value of 
$\lambda$ will finally result in a change to a first order phase transition.
The appearance of these first order phase transitions is a natural 
consequence of 
adding 
the dimenson-6 opeartor, $\left(\varphi^\dagger\varphi\right)^3$, which can lead to multiple
minima of the potential with non-vanishing $vev$.

For $\lambda_6=0.001$ which is shown in fig.~\ref{fig:vev_vs_kappa_l6_0.001}, 
the simulation data and the analytical results from both versions of the 
effective potential agree quite well. 
The results from $U_2$, eq.~\eqref{eq:CEP_with_phi_6_withFullBosDet}, actually coincide with the 
simulation data on a quantitative level as long as the transition is of second order. 
The effective potential $U_1$ reproduces the behaviour of the simulation data qualitatively. 
However, the exact numerical results for the $vev$ differ and the phase transitions are shifted to larger absolute values of $\lambda$.

For $\lambda_6=0.1$ which is shown in fig.~\ref{fig:vev_vs_kappa_l6_0.1}, the effective potential 
$U_1$ shows qualitative agreement with the simulations. The effective potential 
$U_2$ fails to describe the numerical data and the 1-loop evaluation of the 
CEP seems not to be sufficient. 

\begin{figure}[htb]
\centering
\subfloat[$\lambda=-0.0085$]{\includegraphics[width=0.5\linewidth]
{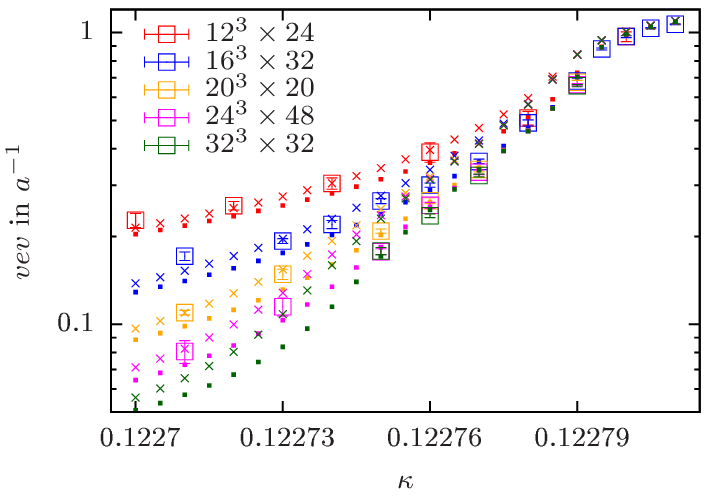}\label{fig:Vdep_l6_0.001_l_-0.0085}}
\subfloat[$\lambda=-0.0088$]{\includegraphics[width=0.5\linewidth]
{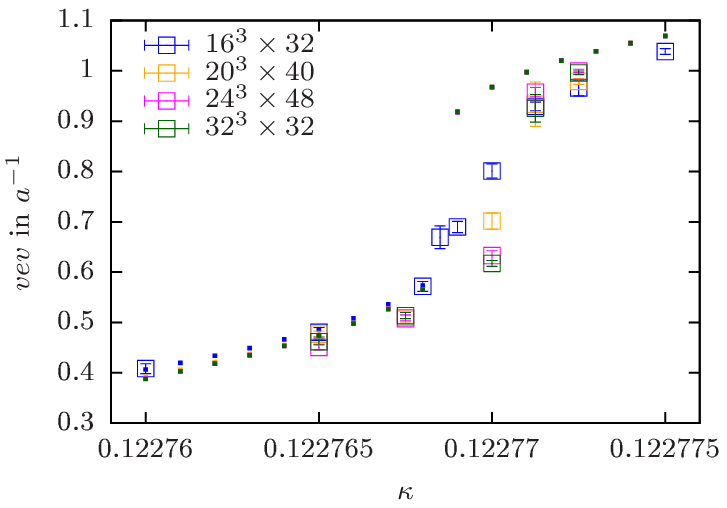}\label{fig:Vdep_l6_0.001_l_-0.0088}}
\caption{
We show the finite volume effects of the phase structure scan for $\lambda_6=0.001$. 
The plot on the left hand side shows data  for $\lambda=-0.0085$ 
where the simulations (open boxes) indicate a second order phase transition.
The plot on the right hand side shows results for $\lambda=-0.0088$, where
the transition is first order. In addition to the simulation data we show the 
data obtained from $U_2$ eq.~\eqref{eq:CEP_with_phi_6_withFullBosDet} (dots) for both 
and from $U_1$ eq.~\eqref {eq:CEP_with_phi_6} (crosses) for the left plot. 
}
\label{fig:Vdep_l6_0.001}
\end{figure}

The results discussed above are obtained on a relatively small lattice of size $16^3 \times 32$. 
To verify the order of the phase transitions, 
simulations and analytical calculations on significantly larger lattices are necessary. 
In figure~\ref{fig:Vdep_l6_0.001}, we show results for the $vev$ as a function of $\kappa$ on various volumes. 
The parameters are chosen in a region where the small volume data indicate 
a second order transition, fig. \ref{fig:Vdep_l6_0.001_l_-0.0085}, and a first order 
transition, fig.~\ref{fig:Vdep_l6_0.001_l_-0.0088}. 
In addition, we compare the simulation data to the analytical results from eq.~\eqref{eq:CEP_with_phi_6} and eq.~\eqref{eq:CEP_with_phi_6_withFullBosDet}. 

%%%%%%%%%%%%%%%%%%%%%%%%%%%%%%%%%%%%%%%%%%%%%%%%%%%%%%%%%%%%%%%%%%%%%%%%%%%%%%%%%%%%%%
%%   vev vs kappa l6=0.001 l=-0.0085, various volumes and l=-0.0088
%%%%%%%%%%%%%%%%%%%%%%%%%%%%%%%%%%%%%%%%%%%%%%%%%%%%%%%%%%%%%%%%%%%%%%%%%%%%%%%%%%%%%
%%%%%%%%%%%%%%%%%%%%%%%%%%%%%%%%%%%
%%%%%%%%%%%%%%%%%%%%%%%%%%%%%%%%%%%

As it is shown in fig.~\ref{fig:Vdep_l6_0.001_l_-0.0085}, the larger volume data confirm the second order nature of the phase transition. 
Furthermore, the finite volume dependence of the second order transition is very well described by both versions of the effective potential.

In fig.~\ref{fig:Vdep_l6_0.001_l_-0.0088}, we show the $vev$ obtained from the 
effective potential $U_2$ and from our non-perturbative lattice simulations on various volumes. 
Both methods give compatible results on a qualitative level and 
just the exact position of the phase transition is slightly altered.
The jump in the $vev$ indicates strongly the existence of a 
first order phase transition at a $\kappa_{\rm trans}\approx 0.12277$.
For these parameter choices, finite size effects are very small. 
In particular, for $\kappa \lesssim \kappa_{\rm trans}$ the $vev$ stays non-zero. 
This means that the first order transition occurs between two minima of the potential with non-zero $vev$. 
Hence, this transition must occur between two broken phases.

\begin{figure}[htb]
\centering
\subfloat[trajectories]{\includegraphics[width=0.5\linewidth]
{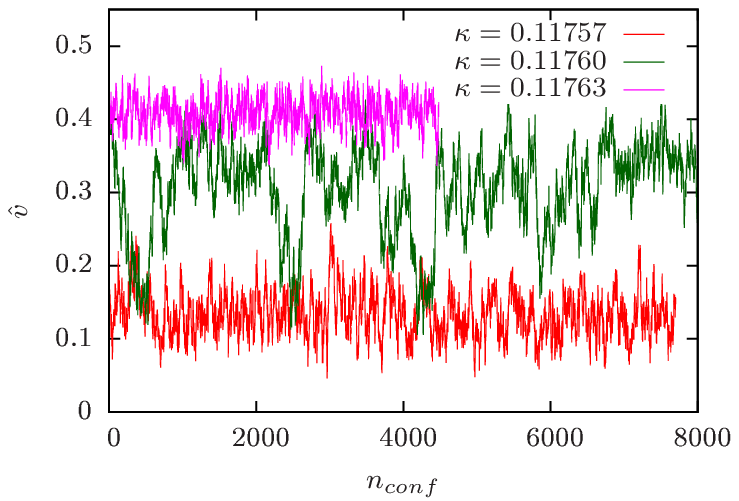}\label{fig:1stOrderPT_simulation_sim}}
\subfloat[CEP from simulations]{\includegraphics[width=0.5\linewidth]
{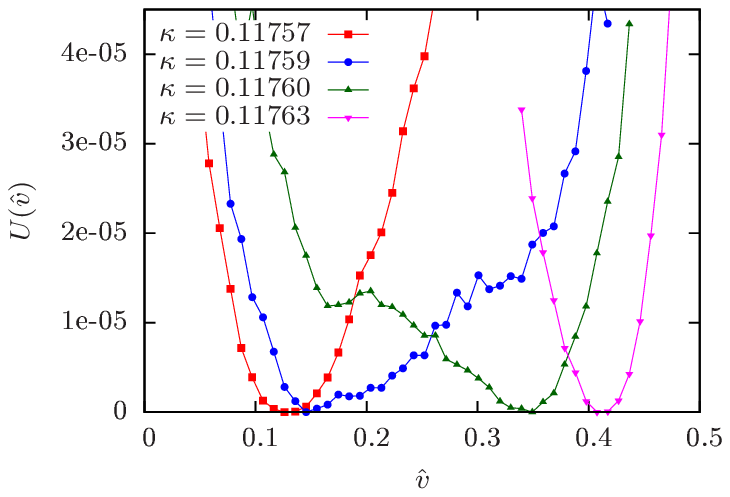}\label{fig:1stOrderPT_simulation_CEP}}
\caption{
The left plot shows the trajectories for ensembles generated around the first 
order phase transition generated on $16 \times 32$ lattices. The 
data correspond to $\lambda_6=0.1$ and $\lambda=-0.38$. The right plot shows the 
corresponding CEP as it was obtained by taking the logarithm of the histograms of the magnetization.
The lines in (b) just serve to guide the eye.
}
\label{fig:1stOrderPT_simulation}
\end{figure}

Close to the point where $\kappa\approx\kappa_{\rm trans}$, tunneling events occur
between the two minima in the simulations and hence the
lattice simulation data may not agree with the results from the effective potential. This stems from the fact 
that the CEP gives only solutions at one of the minima and thus cannot  
take into account tunneling effects. 
In fig.~\ref{fig:1stOrderPT_simulation_sim}
we show the Monte Carlo time history of $\hat v$ at different values of $\kappa$ 
which clearly shows tunneling events.
While for $\kappa=0.11757$ and $\kappa=0.11763$ $\hat v$ fluctuates around the mean value of 
$vev \approx 0.15$ and $vev\approx 0.40$, respectively, for 
$\kappa=0.11760$ tunneling events between these two values appear, 
typical for a first order phase transition.

From the histogram of $\hat v$ with an appropriate binning size, we can construct an effective 
potential from the simulation data. 
This is shown in fig.~\ref{fig:1stOrderPT_simulation_CEP}. 
It is demonstrated nicely 
how the absolute minimum at around $\hat{v}\approx 0.15$ abruptly 
jumps to $\hat{v}\approx 0.35$. Such a behaviour is typical for  
a first order transition.

%%%%%%%%%%%%%%%%%%%%%%%%%%%%%%%%%%%%%%%%%%%%%%%%%%%%%%%%%%%%%%%%%%%%%%%%%%%%%%%%%%%%%%
%%  1st order transition from simulations, trajectory and CEP
%%%%%%%%%%%%%%%%%%%%%%%%%%%%%%%%%%%%%%%%%%%%%%%%%%%%%%%%%%%%%%%%%%%%%%%%%%%%%%%%%%%%%
%%%%%%%%%%%%%%%%%%%%%%%%%%%%%%%%%%%
%%%%%%%%%%%%%%%%%%%%%%%%%%%%%%%%%%%

%%%%%%%%%%%%%%%%%%%%%%%%%%%%%%%%%%%%%%%%%%%%%%%%%%%%%%%%%%%%%%%%%%%%%%%%%%%%%%%%%%%%%%
%%  2nd and 1st order from CEP 
%%%%%%%%%%%%%%%%%%%%%%%%%%%%%%%%%%%%%%%%%%%%%%%%%%%%%%%%%%%%%%%%%%%%%%%%%%%%%%%%%%%%%
\begin{figure}[htb]
\centering
\subfloat[$\lambda_6=0.001, \lambda=-0.0088$]{\includegraphics[width=0.5\linewidth]
{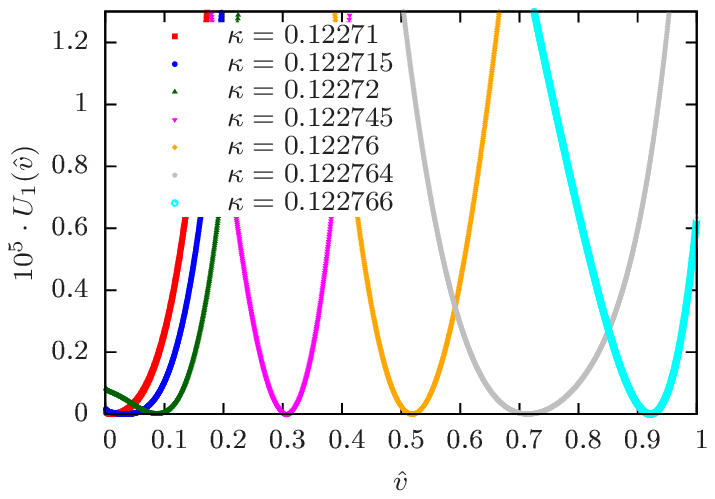}\label{fig:potential_scan_l6_0.001_l_-0.0088}}
\subfloat[$\lambda_6=0.001, \lambda=-0.0089$]{\includegraphics[width=0.5\linewidth]
{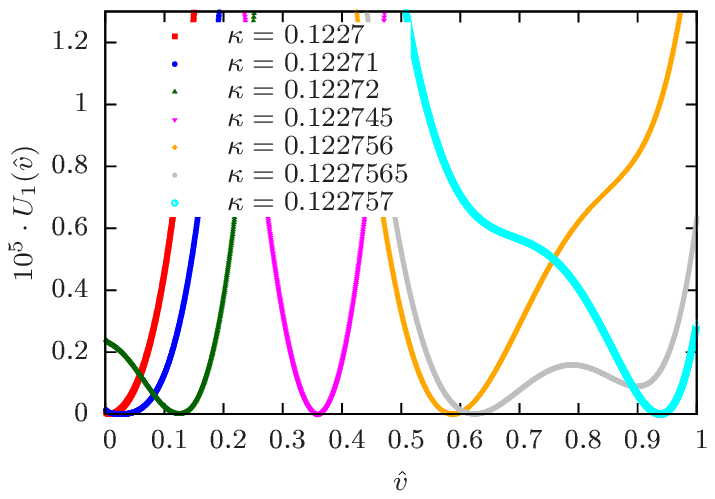}\label{fig:potential_scan_l6_0.001_l_-0.0089}}
\caption{
Here we show the CEP $U_1$, eq.~\eqref{eq:CEP_with_phi_6}, for
fixed $\lambda_6=0.001$ and various $\kappa$ values around the phase transition.
The left plot ($\lambda=-0.0088$) shows a second order phase transition 
for $\kappa\approx 0.122715$. Note that the effective 
potential at $\kappa\approx 0.122764$ actually corresponds 
to a crossover transition, see the discussion in the text 
and fig.~\ref{fig:CEP_Vdep_mass_and_vev}. The right hand plot ($\lambda=-0.0089$) also shows a second order 
transition at $\kappa\approx 0.12271$ and a first order transition $\kappa \approx 0.1227565$
}
\label{fig:potential_scan_l6_0.001}
\end{figure}
%%%%%%%%%%%%%%%%%%%%%%%%%%%%%%%%%%%
%%%%%%%%%%%%%%%%%%%%%%%%%%%%%%%%%%%

Given the fact that for small values of $\lambda_6$ the effective potentials    
describe the simulation data on a quantitative level, it can be utilized 
to investigate the behaviour of the $vev$ further. 
Due to the  wider range of 
applicability we restrict ourselves in the following discussion 
to the potential $U_1$, eq.~\eqref{eq:CEP_with_phi_6}.

We plot the behaviour of the effective potential as a function of $\kappa$ 
in figure~\ref{fig:potential_scan_l6_0.001} for a fixed value of $\lambda_6=0.001$.
In fig.~\ref{fig:potential_scan_l6_0.001_l_-0.0088}, the behaviour 
of the effective potential shows a second order phase transition: the minimum 
moves from a zero to a non-zero value in a smooth way, indicating the second
order nature of the transition. 

However, when $\lambda$ is slightly changed to $\lambda=-0.0089$ we observe, 
in addition to a second order transition at $\kappa\approx 0.12271$, a 
phase transition from one non-zero value of the $vev$ 
to another non-zero value of the $vev$ at large $\kappa$-values, as shown in fig.~\ref{fig:potential_scan_l6_0.001_l_-0.0089}.
This transition happens through a double well potential which is almost realized at $\kappa=0.1227565$. 

\begin{figure}[htb]
\centering
\subfloat[$\lambda=-0.007$]{\includegraphics[width=0.31\linewidth]
{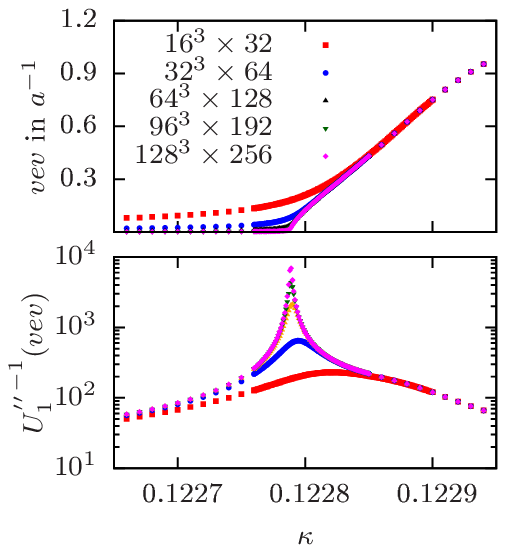}\label{fig:CEP_Vdep_mass_and_vev_l_-0.007}}
\subfloat[$\lambda=-0.0085$]{\includegraphics[width=0.31\linewidth]
{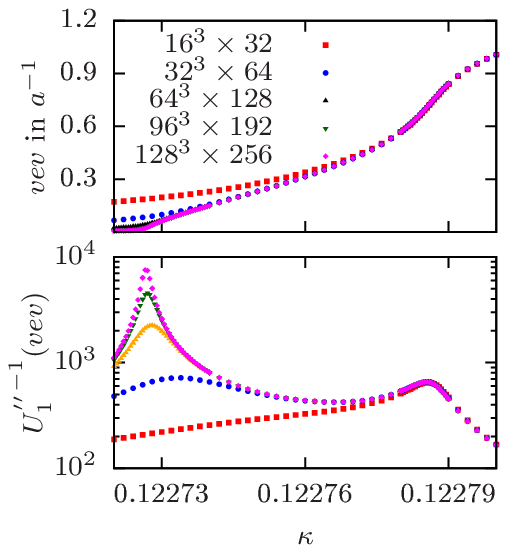}\label{fig:CEP_Vdep_mass_and_vev_l_-0.0085}}
\subfloat[$\lambda=-0.009$]{\includegraphics[width=0.31\linewidth]
{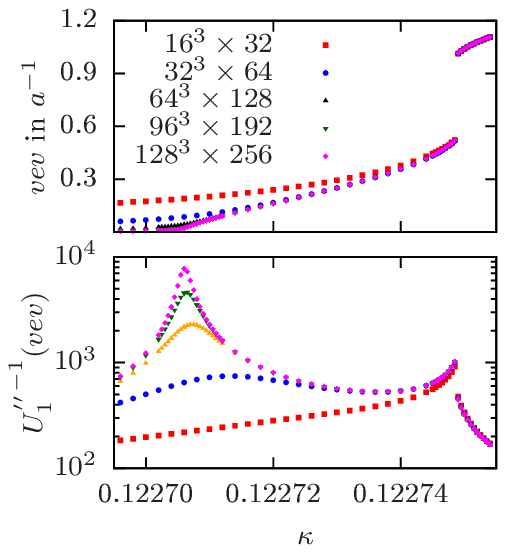}\label{fig:CEP_Vdep_mass_and_vev_l_-0.009}}
\caption{Here the volume dependence of the location of the minimum of the CEP $U_1$, i.e. the $vev$ (upper plots) 
and its inverse curvature in the minimum as a measurement for the magnetic 
susceptibility (lower plots) are shown 
as a function of $\kappa$ for $\lambda_6=0.001$ and a set of $\lambda$-values. 
}
\label{fig:CEP_Vdep_mass_and_vev}
\end{figure}

To determine the location of a second order transition in the CEP, we investigate the curvature of the potential at its minimum, 
$U^{''}(vev)$. The curvature of the potential in its minimum is 
related to the susceptibility $\chi$ of the magnetization, $\chi\propto 1/U^{''}(vev)$,  and is therefore minimal at the 
location of the second order transition. The susceptibility at the phase transition diverges
when the volume goes to infinity corresponding to $U^{''}(vev)$ going to zero. 
To study this finite size effect, we investigate the behaviour of the $vev$ and
the inverse curvature of the potential for volumes up to $128^3 \times 256$. Some example plots are shown in 
fig.~\ref{fig:CEP_Vdep_mass_and_vev} where we plot 
$1/U^{''}(vev)$ as a measure of the magnetic susceptibility. 
In fig.~\ref{fig:CEP_Vdep_mass_and_vev_l_-0.007} the typical behaviour for a second order 
transition is apparent for $\lambda=-0.007$. For $\lambda=-0.0085$ (fig.~\ref{fig:CEP_Vdep_mass_and_vev_l_-0.0085}) 
a second maximum in the inverse curvature of the potential is visible. 
This second maximum is volume independent and indicates a crossover transition 
in the broken phase. 
In fig.~\ref{fig:CEP_Vdep_mass_and_vev_l_-0.009} the second transition 
at $\kappa=0.12275$ has turned into a first order one, while the second 
order 
transition between the symmetric and broken phase is still present at
smaller values of $\kappa$.

%%%%%%%%%%%%%%%%%%%%%%%%%%%%%%%%%%%%%%%%%%%%%%%%%%%%%%%%%%%%%%%%%%%%%%%%%%%%%%%%%%%%%%
%%   Vdep of vev and m_h in CEP l6 0.001
%%%%%%%%%%%%%%%%%%%%%%%%%%%%%%%%%%%%%%%%%%%%%%%%%%%%%%%%%%%%%%%%%%%%%%%%%%%%%%%%%%%%%
%%%%%%%%%%%%%%%%%%%%%%%%%%%%%%%%%%%
%%%%%%%%%%%%%%%%%%%%%%%%%%%%%%%%%%%

\begin{figure}[htb]
\centering
\subfloat[$\lambda_6=0.001$]{\includegraphics[width=0.5\linewidth]
{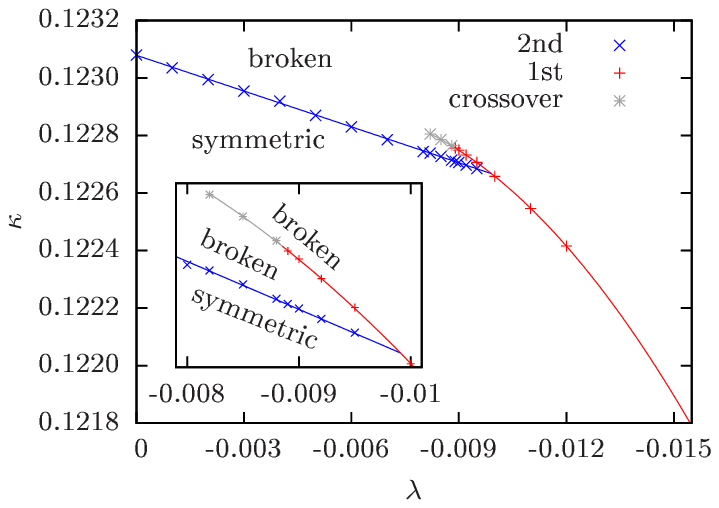}\label{fig:phaseStructure_l6_0.001}}
\subfloat[$\lambda_6=0.1$]{\includegraphics[width=0.5\linewidth]
{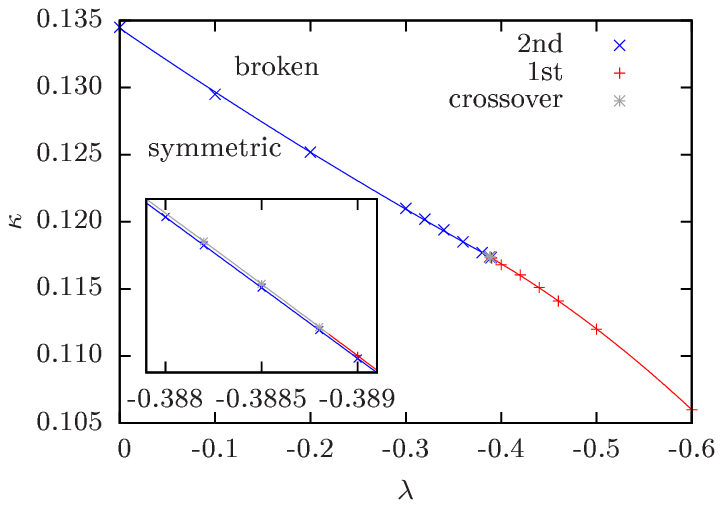}\label{fig:phaseStructure_l6_0.1}}
\caption{
Phase structure obtained from the CEP $U_1$ \eqref{eq:CEP_with_phi_6}. 
There are two phases - a broken and a symmetric one - separated by lines of 
first and second order phase transitions. Furthermore there is a small region in 
parameter space, where a first order transition between two broken phases exists for
$\lambda_6=0.001$ and $\lambda_6=0.1$. The lines between 
the data points are just to guide the eye.
}
\label{fig:phaseStructure}
\end{figure}

Our results for the phase structure computed within the framework of the CEP are summarized in fig.~\ref{fig:phaseStructure} for both $\lambda_6$ values. 
For $\lambda_6=0.001$ we clearly observe a second order phase transition at small absolute values of $\lambda$. 
At intermediate absolute values of $\lambda$ an additional crossover transition sets in within the broken phase. 
This crossover turns into 
a first order phase transition around $\lambda \approx -0.0089$.
The second order transition still exists at this point separating the broken and symmetric phases.
Around $\lambda \approx-0.0098$ and $\kappa \approx 0.12267$ the line of second order transition 
runs into the line of first order transition.
From that point on only the first order transition remains separating the symmetric and broken phases.

%%%%%%%%%%%%%%%%%%%%%%%%%%%%%%%%%%%%%%%%%%%%%%%%%%%%%%%%%%%%%%%%%%%%%%%%%%%%%%%%%%%%%%
%%   phase structure kappa/lambda plane from CEP. lambda_6=0.001 and 0.1
%%%%%%%%%%%%%%%%%%%%%%%%%%%%%%%%%%%%%%%%%%%%%%%%%%%%%%%%%%%%%%%%%%%%%%%%%%%%%%%%%%%%%
%%%%%%%%%%%%%%%%%%%%%%%%%%%%%%%%%%%
%%%%%%%%%%%%%%%%%%%%%%%%%%%%%%%%%%%

For $\lambda_6=0.1$ the general behaviour is very similar. However, the region in parameter space where the additional 
transitions between two broken phases occur is extremely narrow, see the inlet 
in fig.~\ref{fig:phaseStructure_l6_0.1}. 
In fact, the region is so narrow that it is well possible
that in infinite volume only a single transition line
exists with second order transitions for larger and first order 
transitions for smaller quartic couplings. 

\begin{figure}[htb]
\centering
\subfloat[$\lambda_6=0.001$]{\includegraphics[width=0.5\linewidth]
{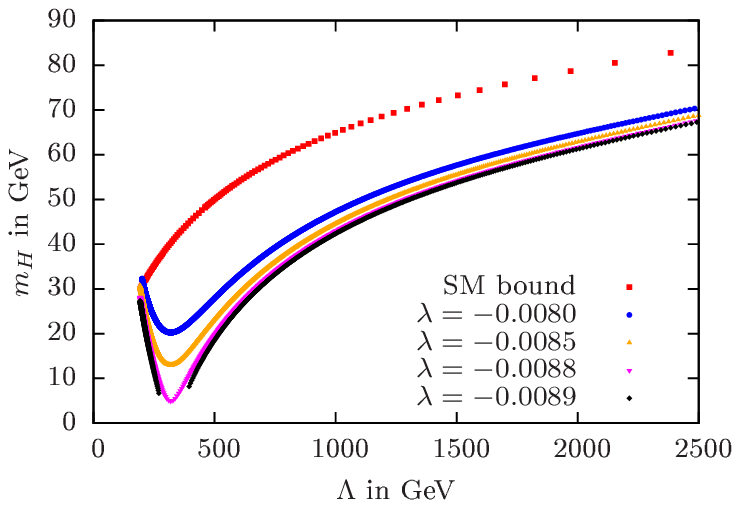}\label{fig:mass_vs_cutoff_l6_0.001}}
\subfloat[$\lambda_6=0.1$]{\includegraphics[width=0.5\linewidth]
{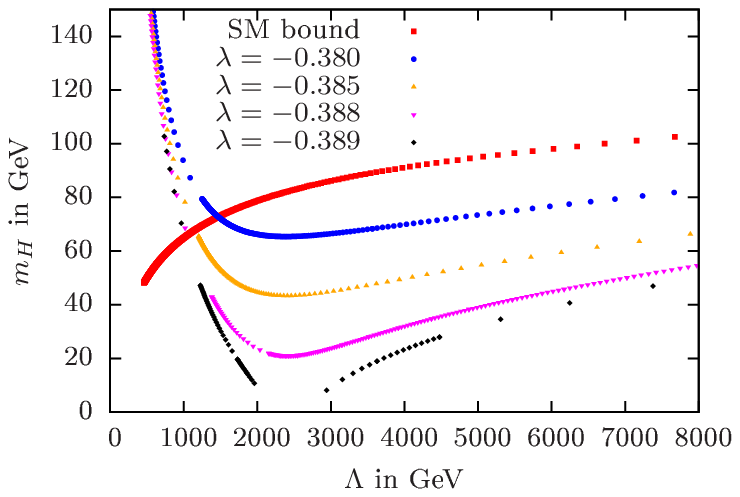}\label{fig:mass_vs_cutoff_l6_0.1}}
\caption{Shown is the cut-off dependence of the Higgs boson mass obtained from the 
CEP according to eq.~\eqref{eq:vev_and_cutoff_CEP} for $\lambda=0.001$ on a $64^3\times128$-lattice (left) and
$\lambda=0.1$ on a $192^3 \times 384$ (right). In both plots we also show the standard model lower mass bound ($\lambda_6=\lambda=0$).
}
\label{fig:mass_vs_cutoff}
\end{figure}

With the CEP the Higgs boson mass can also be obtained from eq.~\eqref{eq:mhSquared_from_CEP}. 
In figure~\ref{fig:mass_vs_cutoff}
we show some first results for the cut-off dependence of the Higgs boson mass obtained by the CEP 
$U_1$ for a series of $\lambda$ values around the region where the first order transitions
appear.
For $\lambda_6=0.001$ 
we observe, see fig.~\ref{fig:mass_vs_cutoff_l6_0.001}, that for the range of cut-off values considered 
here, the Higgs boson mass can be lowered compared to the lower Higgs boson mass 
for vanishing self couplings $\lambda$ and $\lambda_6$ as was also
found in ref.~\cite{Gies:2013fua}. 

Inspecting, however, fig.~\ref{fig:mass_vs_cutoff_l6_0.1}
we find that for $\lambda_6=0.1$ and for small cut-off values, 
the Higgs boson mass 
is significantly larger than the lower bound at vanishing $\lambda$ and $\lambda_6$. 
Note that  
$m_{\rm H}/\Lambda \approx 0.1$, 
i.e. we are still staying in the scaling region of the model.
The increase of the Higgs boson mass at small cut-off  can be understood 
from the fact that the $\lambda_6(\Phi^\dagger\Phi)^3$ term 
in the action provides a positive contribution to the Higgs boson mass shift, 
dominating the negative contribution from the Yukawa coupling. 
For larger values of the cut-off, the $\lambda_6$ coupling becomes less and less 
relevant and the Yukawa term provides the major contribution to the mass-shift such 
that we eventually find the standard behaviour of the Higgs boson mass as a function of 
the cut-off in fig.~\ref{fig:mass_vs_cutoff_l6_0.1}. 

We plan to investigate the cut-off dependence of the Higgs boson mass 
through non-perturbative numerical simulations in the future. However, if the 
picture of fig.~\ref{fig:mass_vs_cutoff_l6_0.1} is confirmed, this would 
lead to a bound on the values of $\lambda_6$ since the $126\,$GeV Higgs boson 
mass would be in conflict with the cut-off dependent mass at low values of the cut-off.
As a consequence, only rather small values of $\lambda_6 \propto O(0.001)$ would be compatible 
with the $126\,$GeV Higgs boson mass.

%%%%%%%%%%%%%%%%%%%%%%%%%%%%%%%%%%%%%%%%%%%%%%%%%%%%%%%%%%%%%%%%%%%%%%%%%%%%%%%%%%%%%%
%%   mass vs cutoff 
%%%%%%%%%%%%%%%%%%%%%%%%%%%%%%%%%%%%%%%%%%%%%%%%%%%%%%%%%%%%%%%%%%%%%%%%%%%%%%%%%%%%%
%%%%%%%%%%%%%%%%%%%%%%%%%%%%%%%%%%%
%%%%%%%%%%%%%%%%%%%%%%%%%%%%%%%%%%%

\section{Conclusions}

In this letter we focused on the investigation of the phase structure of a chirally invariant 
lattice Higgs-Yukawa model  
including an additional higher dimensional
operator, $(\varphi^{\dagger} \varphi)^3$, with coupling strength $\lambda_6$ in the action. 
For the analysis of such a system we restricted 
ourselves to small values of $\lambda_6$ for now. This allowed us 
to compare our numerically obtained results with analytical 
predictions from the constraint effective potential  
evaluated in the same lattice setup as the numerical simulations were carried through. 

In general, we obtained a very good qualitative and even quantitative
agreement between both approaches leading to the phase 
structure shwon in fig.~\ref{fig:phaseStructure} 
for fixed values of 
$\lambda_6=0.001$ and $\lambda_6=0.1$. 

Fixing $\lambda_6>0$ stabilizes the potential, allowing thus to  
drive the values of $\lambda$ 
more and more negative. For sufficiently small values of $\lambda$ we
observe smooth transitions in 
the magnetization, fully compatible with the second order 
phase transitions observed for $\lambda_6=0$. 
However, from a certain negative value of $\lambda$ on,  
we find 
an additional phase transition 
which can be a crossover or  
first order transition. Indications for these transitions can be detected from the 
behaviour of the magnetization computed both in 
the effective potential and the numerical simulations, see e.g. 
fig.~\ref{fig:potential_scan_l6_0.001_l_-0.0089}. 
Thus, the 
resulting phase diagram in fig.~\ref{fig:phaseStructure} turned out to 
be rather rich 
with  
second and first order phase transition lines when changing $\kappa$. 
We note in passing 
that by
fixing the hopping parameter $\kappa$ and hence the bare Higgs boson 
mass, it is possible, to move to a broken phase by only changing the quartic coupling
of the theory. 

A natural extension of the investigation here would be
the exploration of the phase structure of the 
model at non-zero temperature. 
Our results show that
a simple extension of the Higgs-Yukawa sector of the standard model 
by a $(\phi^{\dagger} \phi)^3$ term leads to first order phase transitions. This might 
open the possibility to generate
a strong enough first order phase transition at a non-zero temperature which is compatible 
with baryogenesis \cite{Cohen:1993nk} even at a value of the Higgs boson mass of $126\,$GeV. 

The constraint effective potential also allows to compute the Higgs boson mass
from the second derivative at its minimum. By fixing the value
of $\lambda_6=0.001$ and driving $\lambda$ more and more negative, we obtain
lower and lower values of the Higgs boson mass and, in particular,
substantially smaller values than obtained for $\lambda_6=0$ at a
comparable value of the cut-off. This finding is fully compatible
with the results of \cite{Gies:2013fua}.
As a criterion to obtain an absolute lower bound for the Higgs
boson mass one may choose the value of the quartic coupling, where
the second order standard model like phase transition turns into a
first order one since in the Higgs-Yukawa sector of the SM itself only 
second order phase transitions occur. 

We have also found that
for larger values of $\lambda_6=0.1$ and at small values of the cut-off the
positive contribution of the $\lambda_6$ term to the Higgs boson mass-shift
leads to significantly enhanced Higgs boson masses.
In fact, we can already exclude
certain values of the quartic and $\lambda_6$ couplings since there the
126GeV Higgs boson mass is in conflict with the lower bounds obtained here.
It will be interesting
to perform a more systematic study of the lower Higgs boson
mass bounds at additional values of $\lambda_6$. By employing also 
numerical simulations this can provide exclusion bounds for the coupling values and hence
for models which lead to an extension of the standard
model with a $(\phi^{\dagger} \phi)^3$ term.
We plan to carry out such investigations in the future.

\label{sec:conc}

\section*{Acknowledgments}
We thank K.~Nagai for ongoing discussions and M. M\"uller-Preussker for his continuous support.
We moreover acknowledge the support of the DFG through the DFG-project {\it Mu932/4-4}, the support from Taiwanese MOST via grant
{\it102-2112-M-009-002-MY3} and the support from the DAAD-MOST exchange programme via {\it project 57054177}. 
The numerical computations have been performed 
at the DESY Zeuthen computer center 
 and on the
{\it SGI system HLRN-II} at the {HLRN Supercomputing Service Berlin-Hannover}.

\bibliographystyle{unsrt}
\bibliography{xyukawarefs.bib}

\end{document}